\documentclass[12pt,preprint]{aastex}
\usepackage{amstext}
\usepackage{amsmath}

\begin{document}

\title{A Simple Test of the External Shock Model for the Prompt
  Emission in Gamma-Ray Bursts}

\author{Enrico Ramirez-Ruiz\altaffilmark{1,2,3}, Jonathan
Granot\altaffilmark{4}} \altaffiltext{1}{Institute for Advanced Study,
Einstein Drive, Princeton, NJ 08540, USA} \altaffiltext{2}{Chandra
Fellow} \altaffiltext{3}{Department of Astronomy and Astrophysics,
University of California, Santa Cruz, CA 95064,
USA}\altaffiltext{4}{KIPAC, Stanford University, P.O. Box 20450, MS
29, Stanford, CA 94309, USA}

\begin{abstract} 
It is demonstrated here that if the prompt GRB emission is produced by
the simplest version of the external shock model, a specific relation
should prevail between the observed duration, isotropic equivalent
energy, and photon peak energy. In essence, this relation arises
because both the burst duration and the typical energy of the emitted
synchrotron photons depend on the same combination of the, usually
poorly constrained, external density at the deceleration radius,
$n_{\rm dec}$, and initial bulk Lorentz factor, $\Gamma_0$. This has
the fortunate consequence of making the relation independent of both
$\Gamma_0$ and $n_{\rm dec}$.  Unless the efficiency of electron
acceleration is very low, synchrotron gamma-rays from the external
shock would fail to meet the current observational constraints for the
vast majority of GRBs, including those with a smooth, single peak
temporal profile. This argues either against an external shock origin
for the prompt emission in GRBs or for changes in our understanding of
the microphysical and radiation processes occurring within the shocked
region.
\end{abstract}

\keywords{gamma-rays: bursts -- hydrodynamics -- ISM: jets and
    outflows}

\section{Introduction}
The simplest version of the standard fireball model for gamma-ray
bursts (GRBs) involves a spherical explosion taking place in a uniform
or a stratified surrounding medium. When an explosion deposits a large
amount of energy into material with a much smaller amount of rest
energy within a compact volume, an ultra-relativistic pair fireball is
formed \citep{CR78,Pac86,Goodman86}. The large pressure of the
explosion causes the fireball to expand, and the thermal energy of the
explosion is transformed into bulk kinetic energy due to strong
adiabatic cooling of the particles in the comoving frame. Because of
the Thomson coupling between the particles and photons, most of the
original explosion energy is eventually carried by the baryons that
were originally mixed into the explosion \citep{SP90}.  This bulk
kinetic energy cannot be efficiently radiated as gamma rays unless it
is converted back to internal energy (i.e. the velocities of the
protons must be re-randomized). This requires shocks, and in order to
tap a reasonable fraction of the total kinetic energy, the shocks must
be (at least mildly) relativistic.

Impact on an external medium would randomize about half of the initial
energy merely by reducing the expansion Lorentz factor by a factor of
$\sim 2$ \citep{rm92}. Alternatively, internal shocks may form within
the outflow: for instance, if the Lorentz factor of the outflow varied
by a factor $>2$, then the shocks that developed when fast material
overtakes slower material would be internally (at least mildly)
relativistic. There is a general consensus that the longer complex
bursts must involve internal shocks, while simple smooth profiles
could arise from an external shock interaction
\citep{SP97a,SP97b,rf00,np02,mkp04,rm01}. The latter would in effect
be the beginning of the afterglow.

An external shock moving into a medium with a smooth density profile
would naturally result in a burst with a simple time-profile. Angular
variations within the outflow might still cause variability in the
light curve, but variations on very small angular scales ($\theta <
\Gamma_0^{-1}$, where $\Gamma_0$ in the initial Lorentz factor) are
required in order to produce the large variability of the prompt GRB
emission \citep{f99,dm99}.  A blobby external medium could produce
significant variability only if the covering factor of blobs is low,
implying modest efficiency. Furthermore, the resulting variability in
the light curve would be small if produced close to or after the
deceleration radius, or if the portion of the ejecta that collides
with a blob is decelerated significantly \citep{NG06}.

The purpose of this {\it Letter} is to demonstrate that if the prompt
emission is produced by the simplest version of the external shock
model, this implies a specific relation between the observed duration,
isotropic equivalent energy (or luminosity), and photon peak energy,
which is apparently incompatible with observations.  This relation is
derived in \S~\ref{sec:ES} and compared to observations in
\S~\ref{sec:obs}. The implications are discussed in \S~\ref{sec:dis}
along with possible caveats.

\section{External Shock Model}
\label{sec:ES}

In the simplest version of the external shock model, the outflow is
approximated by a uniform thin shell. A forward shock is driven into
the external medium by the outflowing ejecta, while the latter is
decelerated by a reverse shock (and/or by $pdV$ work across the
contact discontinuity that separates it from the shocked external
medium).  The dynamics of a spherical shock wave eventually approaches
a self-similar evolution \citep{BM76} which depends only on the
explosion energy $E$ and on the external mass density $\rho_{\rm ext}
= n_{\rm ext} m_p$ (the Lorentz factor depends only on their ratio,
$E/\rho_{\rm ext}$). If the initial GRB outflow is collimated, an
additional parameter -- the jet initial half-opening angle,
$\theta_0$, is required in order to specify the flow. However, for
$\Gamma_0\theta_0 \gg 1$ \citep[as appears to be the case from
afterglow modeling;][]{PK02} the dynamics at early times -- before the
jet break time (as long as $\Gamma > \theta_0^{-1}$) do not
significantly deviate from the spherical case, where the true kinetic
energy $E$ is replaced by its isotropic equivalent value $E_{\rm
iso}$. Therefore, it is still valid to adopt the spherical dynamics
for the prompt emission from the external shock in this case as well.

Most of the energy is transfered to the shocked external medium at the
deceleration radius, $R_{\rm dec}$, where the inertia of the swept-up
external matter starts to produce an appreciable slowing down of the
ejecta. For a given shock dynamics, the luminosity and spectrum of the
emitted radiation are determined by the fractions $\epsilon_B$ and
$\epsilon_e$ of internal energy in the shocked fluid that are carried,
respectively, by the magnetic field and by relativistic electrons, as
well as by the shape of the electron distribution function.

As seen in the rest frame of the downstream fluid, most of the mass
and of the kinetic energy of the incoming upstream fluid is in protons
(or other ions), unless the external medium is highly enriched in
$e^\pm$ pairs. Therefore, a simple isotropization of the velocities of
the upstream particles at the shock transition would give the
electrons only a very small fraction of the total internal energy
($\sim m_e/m_p$). This would imply a very small radiative efficiency,
since the radiation is emitted primarily by electrons. For a
radiatively efficient system, physical processes must therefore
transfer a large fraction of the swept-up energy to the electron
component. The energy of the particles can be further boosted by
diffusive shock acceleration as they scatter repeatedly across the
shock interface, acquiring a power law distribution $dN_e/d\gamma_e
\propto \gamma_e^{-p}$ at $\gamma_e > \gamma_m$. 

The strength of the magnetic field is another major uncertainty. Most
of the required magnetic field must typically be generated in-situ,
presumably through plasma instabilities or turbulent motions, but its
strength has yet to be derived from first principles. The standard
prescription is to assume that the magnetic field energy density $U'_B
= (B')^2/8\pi$ is a fixed fraction $\epsilon_B$ of the downstream
proper internal energy density, $B'=(32 \pi \epsilon_B n_{\rm ext} m_p
c^2 \Gamma^2)^{1/2}$, where primed quantities are measured in the
comoving frame.

The typical (minimal) electron energy is given by
\begin{equation}
\gamma_m = 
\frac{m_p}{m_e}\left(\frac{p-2}{p-1}\right)\frac{\epsilon_e}{\xi_e}\Gamma\ , 
\end{equation}
where $\Gamma$ is the Lorentz factor of the fluid behind the forward
shock, and $\xi_e$ is the number of relativistic electrons (or
positrons) per proton, which for a proton-electron plasma is equal to
the fraction of the electrons that are accelerated to relativistic
energies.\footnote{It is assumed here that all the relativistic
electrons take part in the power law distribution of energies; the
definitions of $\epsilon_e$ and $\xi_e$ would not include possible
additional components, such as a thermal component.}

The peak synchrotron photon energy is given by
\begin{equation} 
E_{\rm p} \approx \Gamma\frac{heB'\gamma_m^2}{2\pi m_e c}
= \frac{42\;{\rm keV}}{(1+z)}g^2\epsilon_B^{1/2}\epsilon_e^2
\xi_e^{-2}n_0^{1/2}\Gamma_2^4\ ,
\end{equation}
where $\Gamma_2 = \Gamma(R_{\rm dec})/100$, $n_0$ is $n_{\rm dec} =
  n_{\rm ext}(R_{\rm dec})$ in units of cm$^{-3}$, and $g =
  6(p-2)/(p-1)$ (where $g = 1$ for $p = 2.2$). For $\rho_{\rm ext} =
  n_{\rm ext}m_p = Ar^{-k}$ (with $k < 3$) we have
\begin{eqnarray} 
R_{\rm dec} &=& 
\left[\frac{(3-k)E_{\rm iso}}{4\pi A c^2 \Gamma_0^2}\right]^{1/(3-k)} = 
\left[\frac{(3-k)E_{\rm iso}}{4\pi n_{\rm dec}m_p c^2 \Gamma_0^2}\right]^{1/3}\ ,
\\
T_{\rm dec} &=& (1+z)\frac{R_{\rm dec}}{a c \Gamma_0^2}\ ,
\end{eqnarray}
where\footnote{Note that the initial Lorentz factor of the outflow can
  be higher than $\Gamma_0$ (if the reverse shock is relativistic).}
  $\Gamma_0 = \Gamma(R_{\rm dec}) = \Gamma_{\rm dec}$, $a = 2a_2$ with
  $a_2 \approx 1$, and
\begin{equation} 
n_{\rm dec} \equiv n_{\rm ext}(R_{\rm dec}) = \frac{A}{m_p} 
R_{\rm dec}^{-k} \ .
\end{equation}
In the external shock model the duration of the GRB is $T_{\rm GRB}
  \sim T_{\rm dec}$, and therefore
\begin{equation}\label{psi0} 
\frac{\epsilon_B^{1/2}\epsilon_e^2}{\xi_e^2} \approx 
\frac{4.3\, a_2^{3/2}\sqrt{1+z}}{g^2\sqrt{3-k}}
\left({E_{\rm p} \over 100\;{\rm keV}} \right)
  \left({T_{\rm GRB} \over 20\;{\rm s}}\right)^{3/2} 
\sqrt{\frac{10^{50}\;{\rm erg}}{E_{\rm iso}}}\ .
\end{equation}
Since $\epsilon_B,\epsilon_e \lesssim 1/3$, we can write
\begin{equation} 
\Psi\approx \frac {67\, a_2^{3/2}\sqrt{1+z}}{g^2\sqrt{3-k}}
\left({E_{\rm p} \over 100\;{\rm keV}} \right)
  \left({T_{\rm GRB} \over 20\;{\rm s}}\right)^{3/2} 
\sqrt{\frac{10^{50}\;{\rm erg}}{E_{\rm iso}}} \lesssim \xi_e^{-2}\ .
\label{psi}
\end{equation}
This relatively simple relation between different observable
quantities arises since $T_{\rm dec} \propto R_{\rm dec}/\Gamma_{\rm
dec}^2 \propto E_{\rm iso}^{1/3}(n_{\rm dec}\Gamma_{\rm
dec}^8)^{-1/3}$ while $E_p \propto \Gamma B'\gamma_m^2 \propto
\epsilon_B^{1/2}(\epsilon_e/\xi_e)^2(n_{\rm dec}\Gamma_{\rm
dec}^8)^{1/2}$, so that both $E_p$ and $T_{\rm GRB} \sim T_{\rm dec}$
depend on $n_{\rm dec}$ and $\Gamma_{\rm dec}$ only through the
combination $n_{\rm dec}\Gamma_{\rm dec}^8$. Therefore, the dependence
on both $n_{\rm dec}$ and $\Gamma_{\rm dec}$ (which are hard to
determine from observations) can be eliminated by taking the
combination $E_p T_{\rm dec}^{3/2} \propto
\epsilon_B^{1/2}(\epsilon_e/\xi_e)^2 E_{\rm iso}^{1/2}$.

The strength of Eq.~\ref{psi} is that it depends mainly on quantities
that can either be directly measured, like the peak photon energy
($E_p$) and the duration of the GRB ($T_{\rm GRB}$), or that can be
reasonably constrained by observations. Here $E_{\rm iso}$ is the
isotropic equivalent kinetic energy of the outflow, which for a
reasonable radiative efficiency, $\epsilon_\gamma \sim 0.5$, is of the
order of the isotropic equivalent energy output in gamma-rays,
$E_{\rm\gamma,iso}$, that is measured directly.

\section{Comparison to Observations}
\label{sec:obs}

In order to compare the limit imposed by Eq.~\ref{psi} with
observations, we use the following observational properties derived by
\citet{Ghirlanda04} and \citet{yuki}: $T_{90}$, $E_p$ and
$E_{\rm\gamma,iso}$. In Fig.~\ref{fig1} we show the distribution of
$\Psi$ as a function of $T_{90}$, $E_p$ and
$E_{\rm\gamma,iso}$. Filled circles are {\it typical} long bursts from
the sample compiled by \citet{Ghirlanda04}, while the empty circles
are the four long GRBs found so far to be spectroscopically associated
with type Ic supernovae \citep{yuki}. Of the latter, three have a
smooth, single peak temporal profile (while GRB~030329 has two
peaks). Only two bursts have $\Psi < 1$ while most bursts (and in
particular those associated spectroscopically with supernovae) have
$\Psi \gg 1$. Fig.~\ref{fig2} shows the maximal value of $\xi_e$ that
is allowed according to Eq.~\ref{psi}, $\xi_{e,{\rm max}} =
\Psi^{-1/2}$.

There are some necessary limitations to our approach. The choice of
$\epsilon_e = \epsilon_B = 1/3$ that has been used in the definition
of $\Psi$ in Eq.~\ref{psi} is conservative. More typical values that
are inferred from afterglow modeling ($\epsilon_e \sim 0.1$ and
$\epsilon_B \sim 0.01$) would result in the values of $\xi_e$ being
smaller by a factor of
$8.0(\epsilon_e/0.1)^{-1}(\epsilon_B/0.01)^{-1/4}$ when compared to
$\xi_{e,{\rm max}}$.  It is also important to note that
$E_{\rm\gamma,iso}$ is used as an estimate for the isotropic
equivalent kinetic energy $E_{\rm iso}$. This would increase the value
of $\xi_{e,{\rm max}}$ by a factor of
$[(1-\epsilon_\gamma)/\epsilon_\gamma]^{1/4}$, where $\epsilon_\gamma$
is the $\gamma$-ray efficiency: $E_{\rm
iso}/E_{\rm\gamma,iso}=(1-\epsilon_\gamma)/\epsilon_\gamma$. However,
even for $\gamma$-ray efficiencies as low as $\epsilon_\gamma \sim
10^{-2}$, $\xi_{e,{\rm max}}$ would only increase by a factor of $\sim
3$.

\section{Discussion}
\label{sec:dis}

It has been shown that the simplest version of the external shock
model implies a relation between different observed quantities of the
GRB (Eq.~\ref{psi}), which can conveniently be expressed in the form
$\xi_e \lesssim \xi_{e,{\rm max}} = \Psi^{-1/2}$, where $\xi_e$ is the
number of accelerated electrons per proton. Naively, for the standard
assumption that $\xi_e \approx 1$, one would expect $\Psi \sim
10^{-2}-10^{-1}$ for typical values of $\epsilon_e \sim 0.1$ and
$10^{-3} \lesssim \epsilon_B \lesssim 0.1$. It is conceivable,
however, that only a small fraction of the electrons participate in
the acceleration process (i.e. $\xi_e \ll 1$).\footnote{It should be
pointed out that in principle even $\xi_e > 1$ is possible, especially
near $R_{\rm dec}$, due to pair enrichment of the ambient medium from
pair production by gamma-ray photons that are scattered on the
external medium \citep{rml02,mrrr01,KP04,B02,TM00}.}

\begin{figure}
\plotone{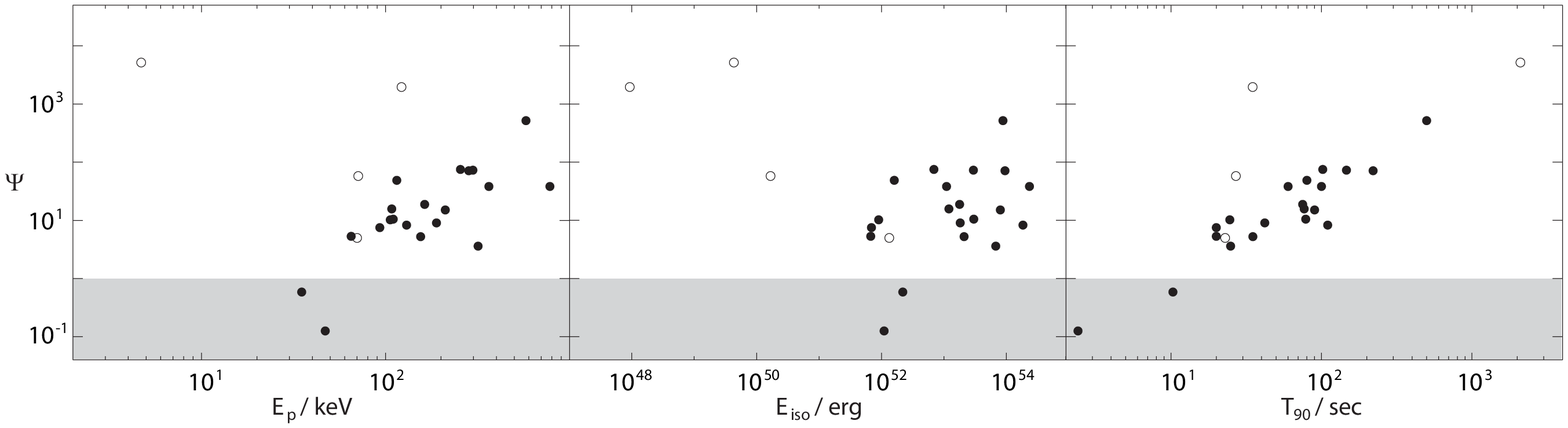}
\caption{{\footnotesize $\Psi$ as a function of $T_{90}$, $E_p$ and
$E_{\rm\gamma,iso}$ for GRBs with established redshifts (black
symbols) from \citet{Ghirlanda04} and for all 4 GRBs with
spectroscopically determined SNe (open symbols) from \citet{yuki}.}}
\label{fig1}
\end{figure}

\begin{figure}
\plotone{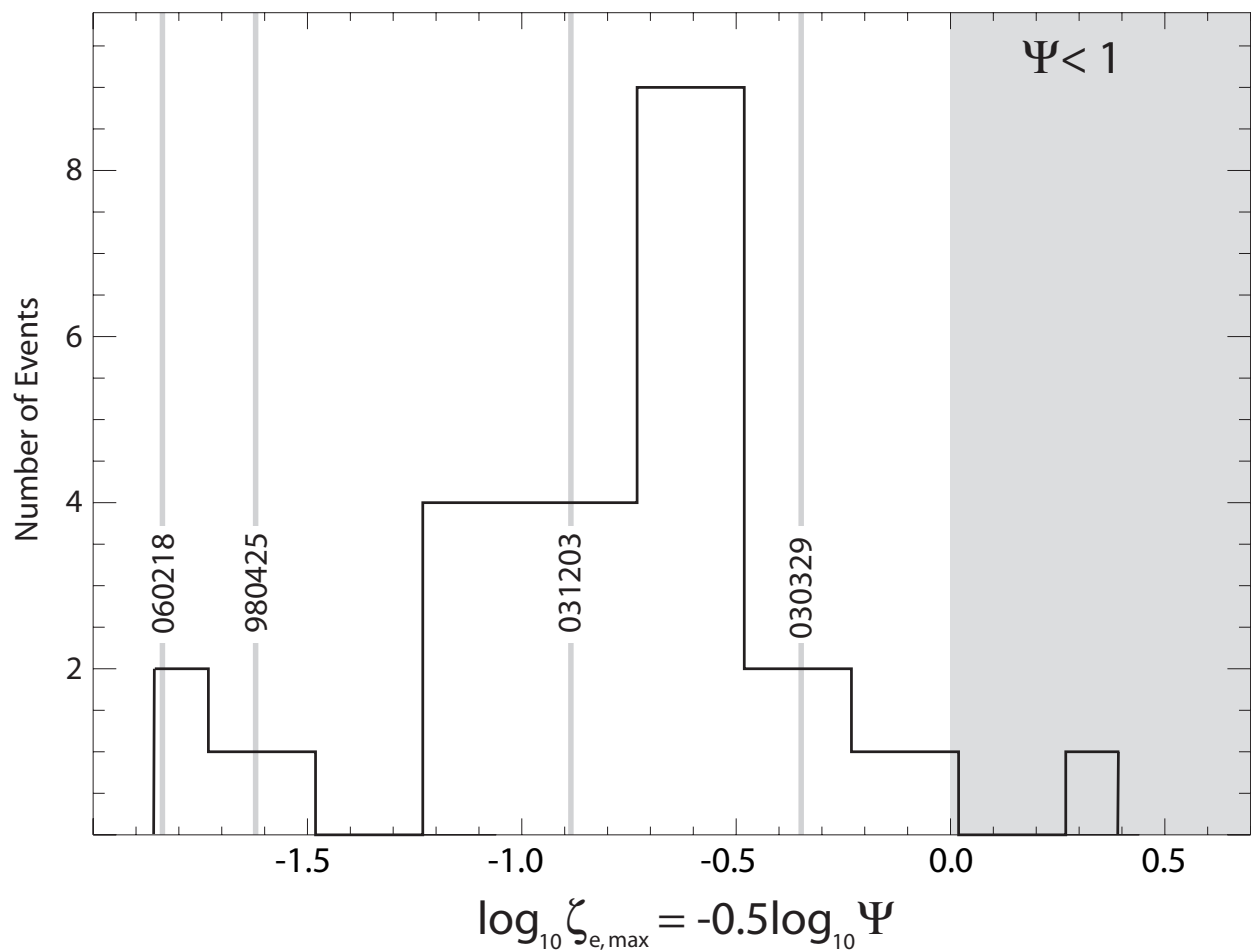}
\caption{{\footnotesize Histogram of the inferred values of
$\xi_{e,{\rm max}}= 1/\sqrt{\Psi}$ for the GRBs in Fig.~\ref{fig1}.}} 

\label{fig2}
\end{figure}

A comparison with observations shows, however, that $\Psi \gg 1$ (and
$\xi_{e,{\rm max}} \ll 1$) for the vast majority of GRBs
(Figs.~\ref{fig1} and \ref{fig2}).  This implies that GRBs could arise
from synchrotron emission in the external shock only if the efficiency
of electron acceleration in relativistic collisionless shocks is very
low ($\xi_e \ll 1$). An external shock origin might still be possible
if the radiation process responsible for the gamma rays is other than
than synchrotron radiation \citep[e.g.,][]{Wang06}. Alternatively, the
prompt gamma-ray emission might arise from completely different
mechanism, such as internal shocks \citep[e.g.][]{rl02}.

It should be noted that afterglow observations already provide
interesting constraints on the efficiency of electron acceleration
\citep{EW05}.  Current observations imply that the characteristic
energy of accelerated electrons is comparable to the proton post-shock
temperature. They also imply that the efficiency $\xi_e$ is similar
for highly relativistic and sub-relativistic shocks and plausibly
suggest that $\xi_e \sim 1$. However, even values of $\xi_e$ as low as
$\sim m_e/m_p$ cannot be ruled out, since currently testable afterglow
predictions remain unchanged for $(E_{\rm iso},n_{\rm ext}) \to
(E_{\rm iso},n_{\rm ext})/\xi_e$ and $(\epsilon_e,\epsilon_B) \to
\xi_e(\epsilon_e,\epsilon_B)$ for any $\xi_e$ in the range $m_e/m_p
\leq \xi_e \leq 1$ \citep{EW05}.

Estimates of the energy in the afterglow shock from late time radio
observations when the flow is only mildly relativistic and starts to
approach spherical symmetry \citep[often called ``radio
calorimetry'';][]{FWK00,BKF04,onp04,Frail05,grl05} typically yield
$E_k \sim 10^{51.5}\;$erg assuming $\xi_e = 1$. However, as noted by
\citet{EW05}, afterglow observations actually constrain $\xi_e E_k$
rather than $E_k$. The true kinetic energies at late times are thus
given by $E_k \sim 10^{52.5}(\xi_e/0.1)^{-1}\;$erg. The initial energy
content of the outflow could be even larger due to early radiative
losses (i.e., during the prompt GRB and early afterglow stages).  It
is difficult to accurately account for the magnitude of such losses,
as they depend on poorly known questions about postshock energy
exchange between protons and electrons. Nevertheless, a lower limit on
the radiated energy is given by $E_\gamma =
E_{\rm\gamma,iso}(1-\cos\theta_0) \approx
E_{\rm\gamma,iso}\theta_0^2/2$ (additional energy may be radiated
outside the observed photon energy range, or during the early
afterglow). Other possible channels of energy loss are the escape of
accelerated non-thermal protons from the blast wave (high energy
cosmic rays) or the production of high energy neutrinos via pion
decay. Any added losses would inevitably lead to a further increase in
the energy requirements.  Therefore, very small values of $\xi_e$
would imply very large energy contents.

Another test of the simple external shock model is provided by a
comparison of the correlation found by \citet{Firmani06} between the
isotropic equivalent luminosity, the burst duration, and the peak
energy, with that predicted by Eq.~\ref{psi0}.  In the cosmological
frame of the GRB this correlation reads $L_{\rm iso} \propto E_{\rm
p}^{1.62\pm0.08}T_{0.45}^{-0.49\pm0.07}$, where $T_{0.45}$ is defined
by \citet{Firmani06} to be the time during which 45\% of the counts
above background are measured (which is expected to scale linearly
with $T_{\rm GRB}=T_{90}$). This is in disagreement with
Eq.~\ref{psi0}, which, for a reasonably small scatter in
$\epsilon_B^{1/2}(\epsilon_e/\xi_e)^2$, gives $L_{\rm iso} \propto
E_{\rm p}^2 T_{\rm GRB}^2$.

It is natural to hope that the values of $\epsilon_B$, $\epsilon_e$,
$p$ and $\xi_e$ are universal, since they are determined by the
microphysics of the collisionless shock.  However, the wide
distribution of $\Psi$ values seen in Fig.~\ref{fig2} suggests
otherwise. That is, in the simplest version of the external shock
model, a large scatter in $\epsilon_B^{1/2}(\epsilon_e/\xi_e)^2$ is
required. The presence of a significant number of non
shock-accelerated electrons in the external shock ($\xi_e \ll 1$)
appears to be more prominent for the sub-sample of bursts found to be
spectroscopically associated with a supernova (Fig.~\ref{fig2}), most
of which have a smooth temporal profile \citep{yuki}. The low values
of $\xi_e$ do not, however, increase the total energy requirements to
unreasonable values for these events as they have rather low values of
$E_{\rm iso}$. Under this interpretation, a wide range of shock
microphysical parameters may be the rule, rather than the exception.

In conclusion, observations of the prompt emission in GRBs with known
redshifts, which are becoming far more accessible in the {\it Swift}
era, can provide an important diagnostic of the external shock
model. Current observational constraints do not allow for efficient
electron acceleration in the external shock, if its synchrotron
emission produces the observed prompt gamma-ray emission. Although
there is no a priori reason to suspect that $\xi_e$ should be large,
$\xi_e \ll 1$ would dramatically increase the total kinetic energy
budget.

\acknowledgments We are especially grateful to Stan Woosley and Eli
Waxman for discussions. This work is supported by IAS and NASA through
a Chandra Postdoctoral Fellowship award PF3-40028 (ERR) and by the
Department of Energy under contract DE-AC03-76SF00515 (JG).


\end{document}